\documentclass[%
 reprint,
nofootinbib,
 amsmath,amssymb,
 aps,
prx,
]{revtex4-2}
\usepackage[english]{babel}
\usepackage{graphicx}
\usepackage{dcolumn}
\usepackage{bm}
\usepackage{lmodern}                       
\usepackage[T1]{fontenc}                   
\usepackage{ae}   
\usepackage[utf8]{inputenc}    
\usepackage{xfrac}  
\usepackage{physics}
\RequirePackage{dsfont}                      

\usepackage[dvipsnames,x11names]{xcolor} 	
\usepackage{comment}    
\usepackage{units}
\usepackage[normalem]{ulem}
\usepackage{blindtext}  
\usepackage{float}  
\usepackage{bbold} 
\usepackage[colorlinks=true,citecolor=OliveGreen,linkcolor=Purple,urlcolor=Magenta]{hyperref} 	
\usepackage{soul}

\begin{document}
\preprint{APS/123-QED}
\title{Distribution of Telecom Entangled Photons through a 7.7 km Antiresonant Hollow-Core Fiber}

\author{Michael Antesberger$^{1, 2, *}$}
\author{Carla M. D. Richter$^{1, 2}$}
\author{Francesco Poletti$^{3}$}
\author{Radan Slavík$^{3}$}
\author{Periklis Petropoulos$^{3}$}
\author{Hannes Hübel$^{4}$}
\author{Alessandro Trenti$^{4, \dagger}$}
\author{Philip Walther$^{1, 2, 5}$}
\author{Lee A. Rozema$^{1, 2}$}

\affiliation{
$^1$ University of Vienna, Faculty of Physics, Vienna Center for Quantum Science and Technology (VCQ), Austria.\\
$^2$ Research Platform for Testing the Quantum Gravity Interface (TURIS), University of Vienna, Austria. \\
$^3$ Optoelectronics Research Centre, University of Southampton, Southampton, United Kingdom. \\
$^4$ AIT Austrian Institute of Technology, Vienna, Austria. \\
$^5$ Christian Doppler Laboratory for Photonic Quantum Computer, University of Vienna, Austria. \\
Correspondence to: $^*$michael.antesberger@univie.ac.at and
$^\dagger$alessandro.trenti@ait.ac.at \\
}


\date{\today}

\begin{abstract}  
State of the art classical and quantum communication rely on standard optical fibers with solid cores to transmit light over long distances. However, recent advances have led to the emergence of antiresonant hollow-core optical fibers (AR-HCFs), which due to the novel fiber geometry, show remarkable optical guiding properties, which are not as limited by the material properties as solid-core fibers. 
In this paper, we explore the transmission of entangled photons through a novel 7.7 km AR-HCF in a laboratory environment at 1550 nm, presenting the first successful demonstration of entanglement distribution via a long AR-HCF. In addition to showing these novel fibers are compatible with long distance quantum communication, we highlight the low latency and low chromatic dispersion intrinsic to AR-HCF, which can increase the secure key rate in time-bin based quantum key distribution protocols.
\end{abstract}

\pacs{Valid PACS appear here}

\maketitle

\vspace{-8mm}
\section{Introduction}
\label{sec:intro}
Over the past few years, quantum technologies such as quantum communication \cite{QKD_review_Gisin} and quantum computing \cite{Deutsch_1998} have made remarkable steps towards maturation. This is currently leading to the emergence of large-scale quantum networks \cite{quantum_internet}, which require quantum communication links between space-like separated nodes \cite{Gisin_quantum_communication}. 
Consequently, flying qubits, encoded in quantum states of light, must be shared between distant parties, using optical fibers. 
Various quantum protocols, including quantum key distribution (QKD) \cite{QKD_review_Gisin}, quantum money \cite{aaronson_quantum_money,schiansky2023demonstration} and quantum coin flipping \cite{quantum_coin_flipping}, rely on low-loss optical links \cite{bozzio2022enhancing}.
In commercially available optical fibers, the lowest propagation loss can be achieved in the relatively narrow telecom C-band ($1530$–$1565$~nm), which is located at an absorption minimum~\cite{buck2004fundamentals_fibers}. In particular, conventional solid-core telecom single mode fibre~(SMF), which is best suited for this wavelength range, offers the minimum absorption at $1550$~nm, while the nearby telecom O-band ($1260$–$1360$~nm) offers zero chromatic dispersion at $\approx1300$ nm \cite{buck2004fundamentals_fibers}.
Moreover, dispersion shifting, via refractive index modification, allows one to shift the zero-dispersion wavelength into the telecom C-band \cite{review_dispersion_shifting}.
Given the optimal performance of solid-core fibers in the C-band, and since quantum protocols are typically very sensitive to loss, most fiber-based quantum communication protocols operate in the C-band.
However, this often brings technical challenges for quantum technologies.
For example, when operating in the C-band one can no longer use cost-effective detection systems, such as Si-Avalanche Photodiodes (Si-APDs) which can achieve quantum efficiencies above $80\%$. Instead, one must use more costly and complex InGaAs-APDs or superconducting nanowire single-photon detector (SNSPD) systems.
In spite of these additional challenges, the low-loss transmission window has led to a concerted effort to modify many quantum technologies away from their natural wavelengths into the C-band.
For instance, quantum dots are a near-perfect single-photon source at $\approx 900$ nm \cite{white_quantumdot}, and achieving the same performance at $1550$ nm remains an open challenge~\cite{Michler_CbandDot,vyvlecka2023robust}.
Similarly, quantum memories, which typically achieve optimal performance in the visible range, are being developed in the C-band \cite{quantummemory_telecomconversion,Pan:20_entanglingmemory}, simply to accommodate solid-core fibers.
Another approach is to frequency shift the emitted photons to the C-band using inefficient nonlinear processes. In fact, this need is so strong, that it has given rise to a new sub-field known as ``quantum frequency conversion'' \cite{deGreve2012Quantumdot,Zaske2012VisibleToTelecom,Morrison2021brightsource,Krutyanskiy2023Multimode}.

\begin{figure*}[ht!]
    \centering
    \includegraphics[width=0.99\linewidth]{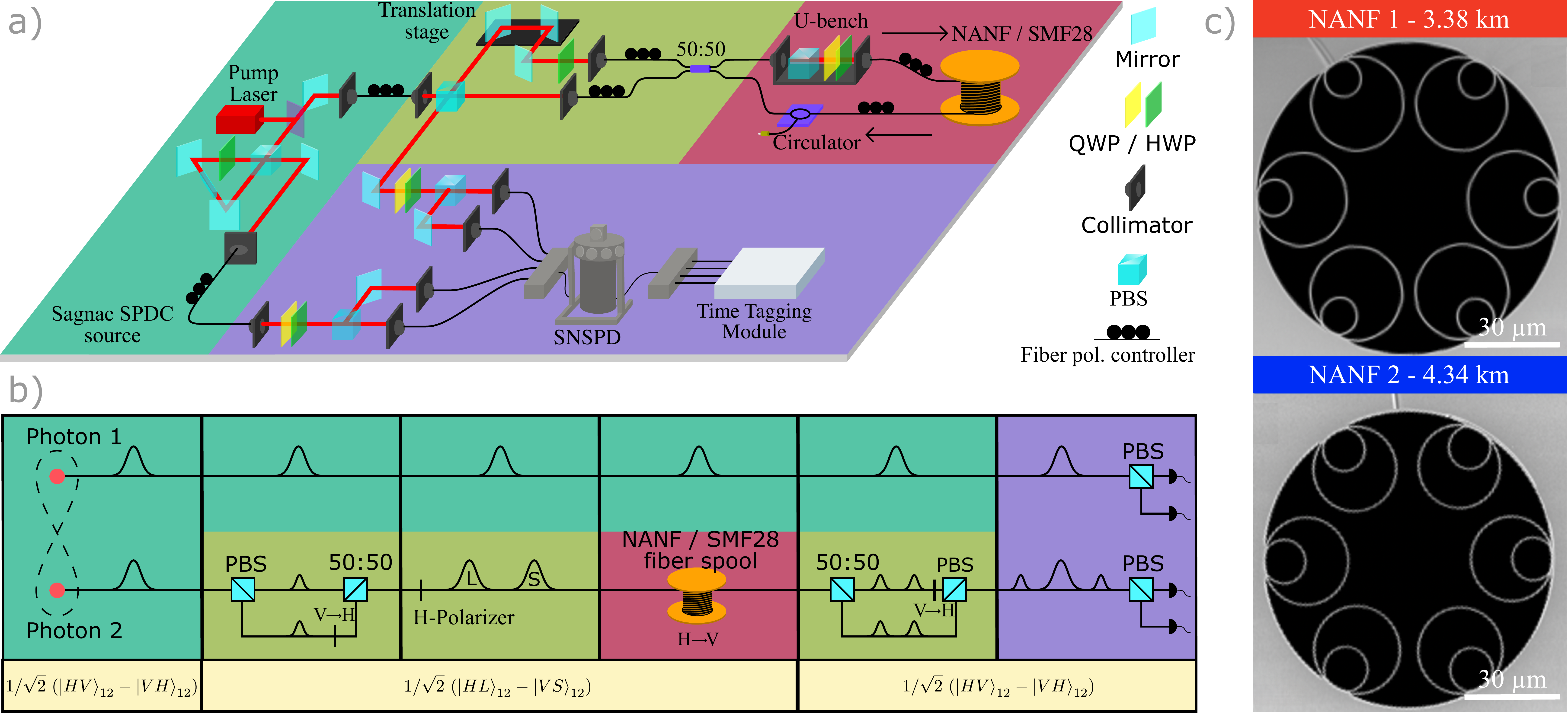}
    \caption{\textbf{Experimental Apparatus:} \textbf{a)} A schematic of the full experimental setup. Panel \textbf{b)} shows a simplified ``unfolded'' setup with color coded panels corresponding to different sections of Panel \textbf{a)}. See the main text for a detailed explanation of each section of the experiment. Panel \textbf{c)} displays a scanning electron microscope image of the cross sections of the two NANFs, which are subsequently spliced together to form the resulting $7.72$ km NANF \cite{Nespola:21_HCF7.72km}.}
    \label{fig::setup}
\end{figure*}

In recent years, antiresonant hollow-core fibers (AR-HCF) \cite{Poletti:14_NANF} 
have emerged as a promising new technology which have several appealing properties that could adress many of these challenges.
AR-HCFs guide light through a central hole surrounded by a microstructure consisting of thin membrains that are in antiresonance with the guided light. 
A scanning electron microscope image of a cross-section of the AR-HCF fibers in our experiment is diplayed in Fig. \ref{fig::setup}c.
The six nested rings along the diameter are the microstructures which guide the light in the central hole.
As most of the light propagates through this hole, light-glass interaction experienced in standard optical fibers with core made of glass is suppressed in AR-HCFs. This brings a range of advantages such as low nonlinearity, low chromatic dispersion, and ability to guide with low loss even at spectral regions, where the fiber glass material experiences appreciable loss~\cite{NumkamFokoua:23_loss_HCF}.

The low loss obtainable with AR-HCFs over a broad wavelength range could be a key ingredient for future wideband quantum networks to transmit quantum states of light at their natural wavelength ~\cite{NumkamFokoua:23_loss_HCF}.
Moreover, the guiding core of AR-HCF, which can be evacuated or filled with gas, exhibits a refractive index of $n\approx1$ \cite{Poletti_HCFvacuumspeedoflight}, resulting in light transmission at approximately the speed of light in vacuum $c$. This holds tremendous potential to revolutionize both classical and quantum communication by achieving the lowest latency possible.
For classical communication, this has sought after applications in financial trading \cite{trading_lightspeed}, clock synchronization, and meeting the latency requirements in 5G and 6G networks \cite{Liu2022Enabling,Sakr2020Interband}. In fact, there is already a commercial market for AR-HCF, with AR-HCF links being deployed for commercial use---the current record being a $40$ km link in the United Kingdom---and active research is being carried out by commercial organizations \cite{Parkin2021eCPRI}.
From a quantum communication perspective, while one may envision similar latency-based enhancements for quantum key distribution or quantum cryptography in general, latency is more of an immediate issue when establishing remote entanglement between matter-based systems.
This is the goal, for example, in quantum repeater networks.
In these scenarios, entanglement is typically transferred between two stationary systems, such as spins in an nitrogen vacancy center \cite{Hensen2015Loophole} or trapped ion qubits \cite{Krutyanskiy2023Entanglement}, by sending single photons through a fiber link.
Therein, it is well-known that the photon propagation time (sometimes called the link time in this context) is a crucial parameter \cite{Krutyanskiy2023Multimode,jones2016design}.
Although the particulars vary from one repeater protocol to the next, the link time has two major influences.
First, some schemes for entanglement swapping require the matter qubit to retain its coherence for the entire photon propagation time \cite{Krutyanskiy2023Multimode}; thus, latency gains on the order of $10$s of microseconds, such as those we observe here, are very relevant, as they are on the same order of magnitude as typical quantum memory lifetimes.
Thus, the use of AR-HCFs would allow the repeater nodes to be placed further apart or use memories with shorter lifetimes.
Second, in quantum repeater protocols,  the link time directly limits the number of times per second one can attempt to establish entanglement, as one must wait this link time to check if entanglement was successfully generated \cite{jones2016design}.
Because this is a probabilistic process, achieving reasonable entanglement generation rates requires as many repetitions as possible.

In quantum repeater networks, the transmitted photon often encodes a time-bin qubit
\cite{repeaterReview}.
From the repeater point of view, this has two advantages: (1) mapping the photonic time-bin to some matter based systems can lead to less decoherence and longer storage times, and (2) time-bin qubits are a leading candidate for transmitting information through fibers.
However,  dispersion in the fiber currently limits the time-bin spacing and thus the maximal key rate \cite{BoaronSimpleTimeBin2018}.
In this respect, the inherently low chromatic dispersion of AR-HCFs can also provide an advantage over solid core fibers. For example, the AR-HCF we use in this work exhibits a dispersion parameter of approximately $2$~ps/nm·km at $\approx1550$~nm \cite{Nespola:21_HCF7.72km}.
While dispersion shifted fibers are somewhat standard for $1550$ nm they are not readily available at wavelengths in the visible spectrum, therein one would need to use a different dispersion compensation technique at the cost of additional loss or complexity.

In addition to their low latency and low dispersion, it has been proposed that AR-HCFs can surpass the fundamental limit of propagation loss imposed by Rayleigh scattering in solid-core fibers \cite{FundamentalsPhotonics} over a broad wavelength range, and this has been recently experimentally realized for wavelengths at $850$ nm and $1060$ nm \cite{Sakr:21_recordloss850_1060}. Since propagation loss is main limiting factor to key-rates in QKD, it is expected that future lower-loss AR-HCF will soon enable longer-distance QKD.
On top of all of these benefits, AR-HCF has a very low intrinsic nonlinearity, enabling the co-propagation of strong classical light with weak quantum signals \cite{Alia2021Coexistence,Honz_co-propagationHCF}.  It is currently not possible to achieve all of these properties simultaneously in solid core fiber.

Given the multitude of appealing properties of AR-HCFs and the fact there is already an emerging market for them in classical communications, they could serve as an excellent candidate to be the backbone for future quantum networks, supporting a diversity of quantum components operating at their natural wavelengths.
In spite of this promise, to the best of our knowledge, the current record for entanglement distribution through a AR-HCF stands at $36.4$ m \cite{Chen:21_HCFentanglement_36m}.
In \cite{Chen:21_HCFentanglement_36m} a latency advantage of only $55$ ns was observed.
Here we present the first distribution of entangled-photon states through a long ($7.7$~km~\cite{Nespola:21_HCF7.72km}) AR-HCF in a laboratory environment, achieving a latency advantage of $13~\mu$s; which could extend both the distance and rate of entanglement generation between matter-based qubits \cite{Krutyanskiy2023Multimode}.
Our fiber is optimized for transmission in the telecom C-band; however, similar performance could be achieved in the visible spectrum. Although this is significantly shorter than the record for entanglement distribution in SMF of $248$ km \cite{Ursin_248km_entanglement}, the potential for AR-HCFs to achieve lower propagation loss than SMF \cite{Poletti:14_NANF} could allow quantum communications in AR-HCF to surpass this limit in the near future.
Furthermore, we highlight the advantage of low dispersion in AR-HCF compared to dispersion unshifted SMF (Corning SMF28) by studying the quality of our entanglement distribution as a function of the time-bin spacing. We achieve the transmission of high-concurrence entanglement at $1550$ nm in AR-HCF using picosecond-spaced time-bin qubits. 

\section{Experimental Implementation}
\label{sec:experiment}
The particular AR-HCF that we study is a $7.7$-km-long hollow-core nested antiresonant nodeless fiber (NANF) \cite{Poletti:14_NANF}.
Although, industry has begun to fabricate HCFs that are tens of kilometers long, to our knowledge the record length reported in an academic setting is an $11$ km photonic bandgab HCF \cite{Chen2015_11km}.
Our specific NANF was previously investigated in Ref. \cite{Nespola:21_HCF7.72km}. It is composed of two individual NANFs, measuring $3.38$~km (NANF~1) and $4.34$~km (NANF~2), that are spliced together (see Fig.~\ref{fig::setup}c).
We determined the total loss of the NANF to be $8.2$ dB, which includes NANF/NANF and NANF/SMF28 splicings. This measurement was done with classical light.
The average propagation loss of the NANF at a wavelength of $1550$~nm is estimated to be $0.82$~dB/km, which exceeds that of standard telecom SMF28 fiber.
Nevertheless, recent advancements have shown that a 
so-called double nested antiresonant nodeless fiber (DNANF) lowers the propagation loss to $0.174$~dB/km \cite{Jasion:22_recordloss0.174}, which is approximately equal to that of SMF28 fiber.
Moreover, AR-HCFs can even surpass the fundamental loss limit imposed by Rayleigh scattering in solid-core fibers~\cite{NumkamFokoua:23_loss_HCF}.

Our experimental setup is illustrated in Fig. \ref{fig::setup}. In brief, we generate polarization-entangled photon pairs. We then convert one polarization qubit into a time-bin qubit, then transmit that time-bin qubit through a long fiber (either NANF or SMF28), and, finally, we convert the time-bin qubit back into a polarization qubit. 
We stress that although the two photons are generated in a polarization-entangled Bell state, while the photon travels through the fiber the photon pair is in a Bell state where the polarization of one photon is entangled to the time-bin of the other.
We implement this ``conversion protocol'' out of experimental convenience primarily for the following four reasons: (1) We can make use of our well established polarization-entangled photon-pair source to readily generate time-bin entanglement; (2) Time-bin encoding is one of the leading candidates for long-distance QKD \cite{marcikic2004distribution}; (3) By converting back to a polarization qubit, we can easily implement arbitrary measurements on the final two-qubit state, allowing for a complete characterization of the entanglement distribution.
(Performing this measurements directly on time-bin qubits is possible, but it is difficult to precisely calibrate the X and Y basis measurements in practice.)
(4) As we will discuss shortly, converting the time-bin qubit back to a polarization qubit to measure it is required to make our experiment passively stable. I.e. we must use the same beamsplitter generate and recombine the time-bin qubit.
Note that while one can directly generate time-bin entanglement from SPDC \cite{Rossi2008timebinentangled}, this often introduces additional systematic errors related precisely setting and stabilizing the relative phase in the time-bin interferometers\cite{marcikic2004distribution}.

To generate entangled photon pairs at $1550$ nm, we use a Type-II SPDC source in Sagnac configuration (Fig. \ref{fig::setup} green area) \cite{SPDC_Kim_PhysRevA}. The measured spectral full width at half maximum (FWHM) bandwidth of the source is $\Delta \lambda \approx0.859$ nm, and assuming transform limited Gaussian pulses, we calculate the coherence time of the photons to be $\tau_{\text{c}}\approx4.1$ ps (which yields $\tau_{\text{c}}\approx2.4$ when expressed as a standard deviation). Initially, we prepare a 2-qubit polarization-entangled Bell state $\ket{\Psi^-}=\frac{1}{\sqrt{2}}\ket{HV}_{12}-\ket{VH}_{12}$~\cite{nielsen_chuang_2010}. ``Photon 1'' is directly coupled to a quantum state tomography (QST) stage consisting of a quarter- (QWP), a half-waveplate (HWP), and a polarizing beam splitter (PBS). Photons at each output port of the PBS are detected by SNSPDs from Single Quantum (Fig. \ref{fig::setup} blue area). Our detectors have an average detector jitter of $\approx 21$ ps, an average detection efficiency of $87\%$, and a dark count rate~$<100$~Hz.

Meanwhile, ``Photon 2'' is coupled to a ``qubit conversion interferometer'' (Fig. \ref{fig::setup} yellow area), which converts the polarization qubit to a time-bin qubit.
To accomplish this, the photon is first sent to a PBS so that if the photon is horizontally (vertically) polarized, it is transmitted (reflected) to the short (long) path. A HWP at $45^\circ$ in the long path flips the polarization state from $\ket{V}$ to $\ket{H}$. Then the two paths are recombined using a fiber 50/50 beamsplitter (BS). 
We discard the cases wherein Photon~2 exits in the lower arm of the BS using the optical circulator shown in the red section of Fig. \ref{fig::setup}. While this could be avoided using active optical switches \cite{zanin2021fiber,antesberger2023higher}, for our picosecond time-bin spacings this would require GHz switching speeds.
This results in Photon~2 encoding a time-bin qubit in the upper output mode of the BS.
Photon 2's time-bin qubit is entangled with the polarization qubit encoded in Photon 1.
Thus, the Bell state can be written as $\ket{\Psi^-}=\frac{1}{\sqrt{2}}\ket{HL}_{12}-\ket{VS}_{12}$, where $\ket{S}$ ($\ket{L}$) refers to the photon taking the short (long) path of the interferometer. 
Furthermore, we can adjust the time-bin spacing $\Delta t$ by tuning a delay line in the long path.

Photon 2 then proceeds to the red area of Fig. \ref{fig::setup} (propagating clockwise through the loop), and passes a short free-space u-bench hosting a PBS to ensure that both time-bins are horizontally polarized. Up to this point, we use manual fiber polarization controllers to set all fiber-induced polarization transformations to identity. After the u-bench, we send the photon to either the $7.7$ km NANF or a SMF28 fiber of comparable length ($7.8$~km).
Thus, although Photon 2 is generated as a part of a polarization entangled photon pair, as it propagates though the fiber it is a time-bin qubit, which is entangled to the polarization of Photon 1.

After the fiber transmission, the fiber is connected to the circulator, and then to the lower port of the qubit conversion interferometer (back in the yellow area of Fig. \ref{fig::setup}). Now, as the photon traverses the interferometer in reverse direction, the time-bin qubit encoded in Photon 2 is converted back to a polarization encoding. 

In order to convert the time-bin qubit back to a polarization qubit, the long time-bin mode needs to be delayed, after which the two time bins should be recombined on the PBS. However, since we use passive optics, we cannot selectively delay the long time bin. In other words, at the BS we want the long (short) time bin to take the short (long) path, but half of the time the opposite situation occurs.
This leads to three prominent peaks in the photon arrival time in the output path of the PBS (see Fig. \ref{fig::setup}b). The two side peaks correspond to $\ket{S}$ ($\ket{L}$) taking the short (long) path. Whereas the center peak arises from the coherent recombination of $\ket{S}$ and $\ket{L}$. Hence, the polarization in the central peak is now re-entangled with the polarization of Photon 1. Now, back in polarization basis and post-selecting on the central peak, the ideal Bell state can be then written as $\ket{\Psi^-}=\frac{1}{\sqrt{2}}\ket{HV}_{12}-\ket{VH}_{12}$. By using the same conversion interferometer in a loop geometry (Fig. \ref{fig::setup} red area) we can ensure that the recombination of the time-bins is passively phase stable. The time-bin processing scheme employed here is an extension of the approach presented in \cite{antesberger2023higher}, adapted for ultra-short entangled time-bins. In order to ensure that Photon 2 is sent from the first PBS of the conversion interferometer to QST stage (Fig. \ref{fig::setup} blue area) we use fiber polarization controllers before and after the NANF/SMF28 spool together with the QWP and HWP in the u-bench such that Photon 2's polarization is flipped from $\ket{H}$ to $\ket{V}$ as it traverses the loop. Finally, after the PBS in the QST stage Photon 2 is detected using SNSPDs.

\section{Results}
\label{sec:results}
\begin{figure}[h]
    \centering
    \includegraphics[width=\linewidth, trim=0.cm 1.1cm 1.25cm 2.75cm, clip]{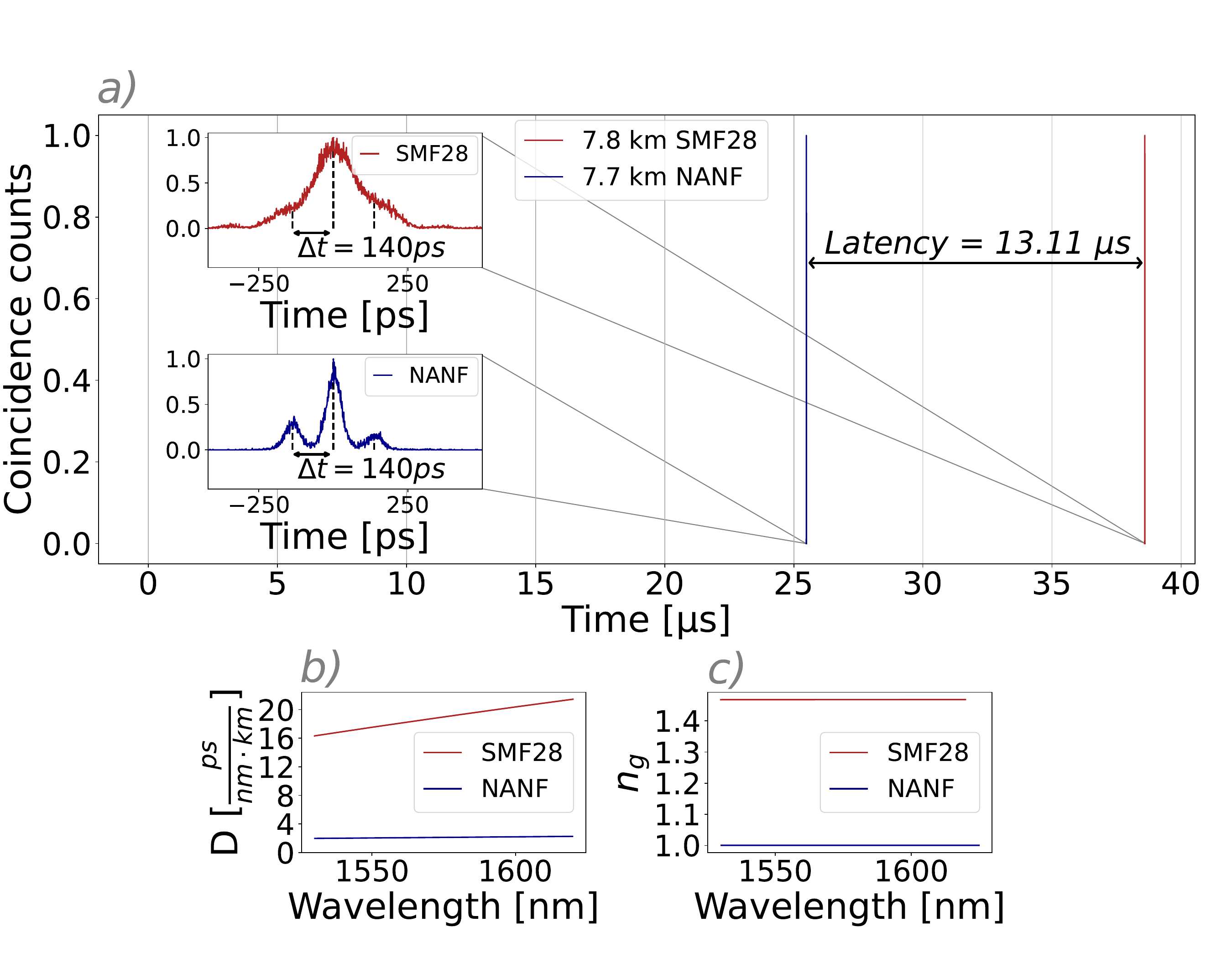}
    \caption{\textbf{Latency Measurements:} \textbf{a)} The normalized arrival-time histogram between Photon 1 and 2 after transmission of the entangled time-bin qubit through either $7.7$ km of NANF (blue) or $7.8$ km SMF28 (red) fiber. The photons arrive $13.11~\mu$s earlier when traversing the NANF. Insets: Zoomed view on both major peaks, showing three peaks discussed in the main text. The effect of dispersion in the SMF28 fiber is evident in the upper inset. \textbf{b)} The dispersion parameter $D$ of both fibers as a function of wavelength. \textbf{c)} The group refractive index of NANF and SMF28 plotted versus wavelength.}
    \label{fig::latency_histogram}
\end{figure}

We will first present the results of our the latency measurements. The group velocity of light in a guided mode ($v_{g} = c/n_g$, where $c$ is the speed of light in vacuum) is determined by $n_g$, the group refractive index of the mode \cite{FundamentalsPhotonics}.
At $1550$ nm NANF has a group refractive index of approximately $n_{\text{HCF}}\approx 1$, while in SMF28 fiber $n_{\text{SMF28}}\approx1.47$ (see Fig. \ref{fig::latency_histogram}c).

\noindent Consequently, $v_{g, \text{NANF}} > v_{g, \text{SMF28}}$. We demonstrate this by recording an arrival-time histogram between the Photon 1 and Photon 2. To do so, Photon 1 is directly detected at the QST stage (see Fig. \ref{fig::setup}), while Photon 2 undergoes the encoding conversion and propagates through either the $7.7$~km NANF or the $7.8$~km SMF28 fiber spool before being detected at the second QST stage. The resulting histogram is illustrated in Fig.~\ref{fig::latency_histogram}a, we observe an average coincidence rate of $233.2$ Hz ($29.4$ Hz) when the photon is transmitted through the SMF28 (NANF). As expected, when Photon 2 traverses the SMF28 solid-core fiber it arrives $\approx13$~$\mu$s later than when it propagates through the NANF, resulting in an approximately $34$~\% lower latency in NANF compared to SMF28. For this measurement we chose a time-bin spacing of $\Delta t=140$~ps. The inset plots in Fig. \ref{fig::latency_histogram}a also show that, due to the larger dispersion of SMF28 compared to NANF (see Fig. \ref{fig::latency_histogram}b), the three prominent peaks from our passive time-bin recombination are almost entirely unresolvable. However, after transmission through NANF, these peaks remain intact. We address this quantitatively later, by varying the time-bin spacing.
Note that the unequal height of the side peaks is caused by imbalanced losses in the two arms of the interferometer. This does not affect the central peak because these detection events arise from instances wherein the photon has taken the short (long) path in the forwards direction and the long (short) path in the reverse direction.
Thus, other than the overall loss, this imbalance does not affect properties of the quantum state, which is only reconstructed from events in the central peak.

\begin{figure}[t]
    \centering
    \includegraphics[width=\linewidth, trim=12.3cm 1.8cm 9.7cm 1.8cm, clip]{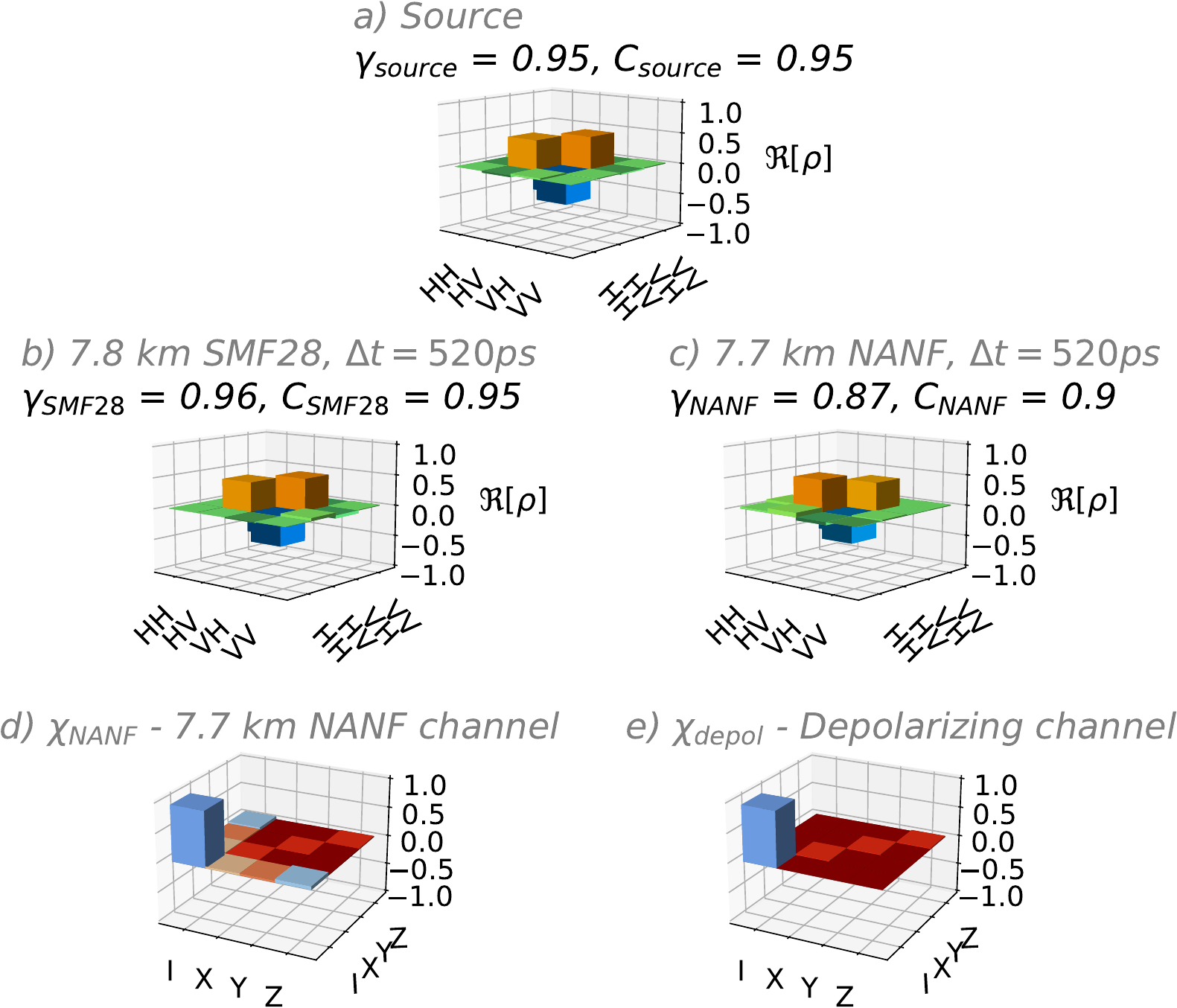}
    \caption{\textbf{Entanglement Distribution.}  The density matrices measured before being converted to a time-bin qubit and transmitted through a long fiber (\textbf{a}), after conversion to a time-bin qubit with $\Delta t=520$ ps and being distributed through (\textbf{b}) the $7.8$ km SMF28 fiber and (\textbf{c}) the $7.7$ km NANF. \textbf{d}) The reconstructed $\chi$-matrix of the NANF. This describes the quantum channel experienced by a polarization state traversing the fiber. \textbf{e}) The theoretical $\chi$-matrix of a purely depolarizing channel, with the depolarization strength pf $p=0.94$ set to match the NANF.}
    \label{fig::2qubit_1qubit_densities}
\end{figure}

To demonstrate the distribution of high-concurrence entanglement through a long NANF, we first characterize the entangled state (nominally a $\ket{\Psi^-}$ state) produced by our source using two-qubit quantum state tomography (QST).
The concurrence $C$ is measure of entanglement, where for a maximally entangled state $C=1$, for a fully separable state $C=0$, and for any non-zero value of $C$ the state is partially entangled \cite{concurrence_PhysRevLett.78.5022}.
We generate entangled photon pairs using the SPDC source depicted in Fig.\ref{fig::setup}. The density matrix $\hat{\rho}_{\text{source}}$, reconstructed by QST, is illustrated in Fig. \ref{fig::2qubit_1qubit_densities}a. It has a purity of $\gamma_{\text{source}}=0.9493\pm0.0008$ and a concurrence of $C_{\text{source}}=0.9482\pm0.0007$. 
Throughout our work, the errors for all results extracted from QST (i.e. the purity and concurrence) are numerically estimated using Monte-Carlo simulations to account for Poisson counting statistics.
We then convert Photon 2's qubit from polarization to time-bin and distribute it through a $7.8$ km SMF28 fiber. 
To ensure that the time-bin qubit is not affected by dispersive pulse broadening we use a large time-bin spacing of $\Delta t = 520$ ps. 
Performing QST after this process confirms that the input state is almost unchanged, with $\gamma_{\text{SMF28}}=0.956\pm0.002$ and $C_{\text{SMF28}}=0.946\pm0.002$ (see the density matrix $\hat{\rho}_{\text{SMF28}}$ in Fig. \ref{fig::2qubit_1qubit_densities}b). 

\begin{figure*}[t]
    \centering
    \includegraphics[width=\linewidth, trim=3.cm 1.8cm 0.1cm 0.9cm, clip]{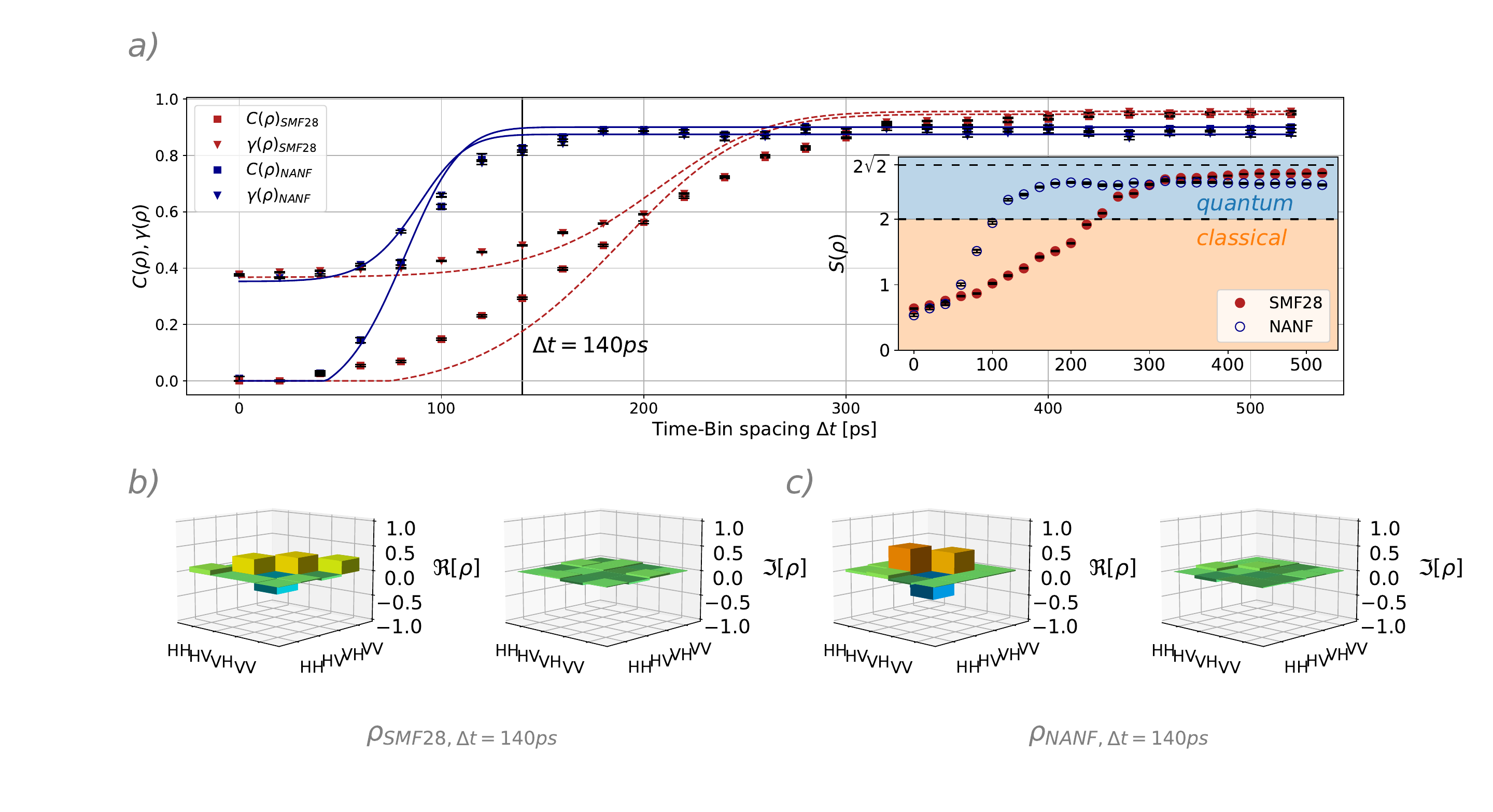}
    \caption{\textbf{Time-Bin Spacing Dependence.} \textbf{a)} The concurrence and purity of the $\ket{\Psi^-}$-Bell state after propagation of the time-bin qubit through either $7.7$ km NANF (blue) or $7.8$ km SMF28 fiber (red) with different time-bin spacings $\Delta t$. The blue triangles (squares) corresponds to the measured purity (concurrence) of the state after NANF transmission, the solid blue lines represents the simulated data. The red markers and dashed lines correspond to data taken through the SMF28 fiber. The inset in panel \textbf{a} illustrates the CHSH inequality's value $S$ as a function of the time-bin spacing of the distributed states. Panel \textbf{b} (\textbf{c}) shows the real and imaginary parts of the reconstructed density matrix after distribution through the SMF28 fiber (NANF), with a time-bin spacing of $\Delta t = 140$ ps.}
    \label{fig::concurrence_purity_VS_dt}
\end{figure*}

We then repeat the experiment with the $7.7$ km NANF. A QST measurement again confirms that we have successfully distributed one photon of an entangled Bell state through the NANF while maintaining a high concurrence. In particular, we find $C_{\text{NANF}}=0.901\pm0.006$ and $\gamma_{\text{NANF}}=0.875\pm0.006$ for $\hat{\rho}_{\text{NANF}}$. The full density matrix is illustrated in Fig. \ref{fig::2qubit_1qubit_densities}c. We do note, however, that  $\hat{\rho}_{\text{NANF}}$ has a reduced concurrence compared to the SMF28 results. This is due to a slight depolarization of the photon. The origin of this depolarization effect is not clear, but it is likely related to polarization mode dispersion in the NANF or inter-modal coupling at the NANF-NANF splice. 
Since this can only be confirmed by destructive measurements, we leave this for future investigation.We further note that early solid-core single-mode fiber suffered similar depolarization from manufacturing defects \cite{burns1983depolarization}, and as fabrication of AR-HCF improves this issue will likely be resolved.
In fact, under certain conditions, AR-HCFs can already be optimized for extremely high-purity polarization transmission \cite{HCF_high_purity}. Since the transmitted qubit is encoded in the time degree of freedom (DOF), this depolarization should not affect the distributed quantum state. However, because we convert the time-bin qubit back to polarization, the converted polarization state is degraded by the depolarization. This issue could be overcome by adding a polarizer in the setup before performing the back-conversion of the qubit to polarization. However, this would be accompanied by higher photon loss. Alternatively, with the use of ultra-fast active optic elements (e.g. phase shifters and optical switches), one could directly perform QST on the time-bin qubit.

To verify the mild depolarization in our NANF we perform ancilla-assisted quantum process tomography on the NANF polarization channel \cite{ancillaProcessTomo_PhysRevLett.90.193601} to reconstruct the underlying quantum process described by a $\chi$-matrix \cite{nielsen_chuang_2010}.
While this measurement could be performed using classical light, the degree of depolarization strongly depends on the spectral shape of the light. Therefore, it is more accurate to measure the depolarization effects directly using the entangled photons source. To perform this process tomography, we directly send Photon 2 through the NANF without converting it to a time-bin qubit, using Photon 1 as the ancilla qubit. The process describing the transmission of Photon 2 through the NANF can then be directly extracted from the same measurements used to do two-photon state tomography.
The NANF-channel is described by $\chi_{\text{NANF}}$, which we plot in Fig. \ref{fig::2qubit_1qubit_densities}d. For comparison, we model a uniform depolarizing channel of the form $\chi_{\text{depol}}(\rho) = p \mathbb{1} \rho \mathbb{1} + \frac{1-p}{3}(X \rho X + Y \rho Y + Z \rho Z)$, where $X$, $Y$ and $Z$ are the three Pauli operators (see Fig. \ref{fig::2qubit_1qubit_densities}e) \cite{nielsen_chuang_2010}. 
We then set the probability of the identity component, $p$, in $\chi_{\text{depol}}$ to match the experimentally reconstructed probability in $\chi_{\text{NANF}}$, which is $p=0.94 \pm 0.02$. We calculate the fidelity between both channels, finding $F(\chi_{\text{NANF}}, \chi_{\text{depol}})=0.988$. The main discrepancy between our simple model and the experimental channel is evident in the slight non-zero off-diagonal elements in $\chi_{\text{NANF}}$, introducing a dependence of the output state purity on the input polarization. 
In other words, $\chi_{\text{depol}}$ describes completely uniform depolarization for all input states, whereas $\chi_{\text{NANF}}$ has a preferred transmission axis.
We numerically estimate the best and worst case state purities to be $0.97$ for approximately vertically-polarized photons and $0.92$ for horizontally-polarized photons.

So far, we have demonstrated the low-latency distribution of entangled photons through NANF at a group velocity $v_{g, \text{NANF}}\approx c$. However, NANFs also naturally possess low chromatic dispersion, which is crucial for distributing ultra-short time-bins over long distances. When the closely-spaced time-bins overlap in the fiber due to dispersion, the quality of the entanglement is compromised.
We thus repeat the entanglement distribution using different time-bin spacings $\Delta t$, from $0$ to $520$ ps with a step size of $20$ ps in both fibers. As the time-bins start to overlap (in practice, when $\Delta t < 6\sigma$), the concurrence and purity of the distributed state decreases. Here, $\sigma$ represents the standard deviation of the coincidence histogram recorded by our detectors and time tagger (Swabian Instruments, TimeTagger Ultra), which is a combination of the width of the photon wavepacket and the timing jitter of our detection system. 

We find that after propagating through a 7.8 km SMF28 fiber the width of the time bin peaks is $\sigma_{\text{SMF28}}\approx54.1\pm0.3$ ps, while after a 7.7 km NANF, the histogram exhibited a smaller standard deviation of $\sigma_{\text{NANF}}\approx23.1\pm0.3$ ps (which is dominated by the detection system jitter). To verify this, we measure the same histogram directly from the source (without a long fiber), observing $\sigma_{\text{source}}\approx21.1\pm0.2$ ps. This confirms that most of the width in our NANF measurements comes from limitations in our detection system.  
{In particular, from the dispersion parameter $D$ and the spectral width of the photons ($\Delta \lambda \approx 0.859$ nm FWHM, or $\Delta \lambda \approx 0.365$ nm in standard deviation),
we can determine the expected pulse broadening after propagating through the respective fibers. Assuming Gaussian wavepackets, the pulses will experience an additional temporal broadening of $\Delta \tau(z) \approx D \Delta \lambda z$, where $z$ represents the length of the dispersive medium \cite{FundamentalsPhotonics}. For SMF28 fiber with $D_{\text{SMF28}} \approx 18$~ps/nm·km, we anticipate a pulse broadening of $\Delta\tau_{\text{SMF28}}(z=7.8~\text{km})~\approx~51.2$~ps, in standard deviation. On the other hand, our NANF exhibits lower dispersion with $D_{\text{NANF}} \approx 2$ ps/nm·km, resulting in a  pulse broadening of only $\Delta\tau_{\text{NANF}}(z=7.72 \text{km}) \approx 5.6$ ps in standard deviation. Note that the specific values of the dispersion parameter $D$ were measured with standard techniques using classical light.
Combining this with our system jitter we expect to observe a width of $\sigma_{\text{SMF28}}' = \sqrt{\Delta\tau_{\text{SMF28}}^2 + \sigma_{\text{source}}^2} \approx 55.4$ ps for SMF28 fiber and $\sigma_{\text{NANF}}' = \sqrt{\Delta\tau_{\text{NANF}}^2 + \sigma_{\text{source}}^2} \approx 21.8$ ps for NANF. 
Both of these values agree fairly well with our measured widths. Furthermore, this suggests that our observed $\sigma_{\text{NANF}}$ is primarily influenced by the system jitter, which means that it should possible achieve even smaller time-bin spacings in NANF.}

Our data showing the dependence of the two-qubit concurrence $C$ and purity $\gamma$ on the time-bin spacing for NANF and dispersion unshifted SMF28 fiber is presented in Fig. \ref{fig::concurrence_purity_VS_dt}a. 
Clearly, NANF outperforms SMF28 fiber as the time-bin spacing $\Delta t$ is reduced.
This is further reflected in the transmitted state's ability to violate the Clauser-Horne-Shimony-Holt (CHSH) inequality \cite{1969_CHSH}, plotted in the inset of inset of Fig. \ref{fig::concurrence_purity_VS_dt}a. 
In SMF28, we find that the concurrence, purity and CHSH values begin to drop at $\Delta t\approx 300$ ps, while, due to the smaller dispersion parameter in NANF, this drop off does not occur until $\Delta t\approx140$ ps in NANF.
To illustrate this effect, we present the density matrices measured with $\Delta t = 140$ ps through SMF28 fiber ($\rho_{\text{SMF28}, \Delta t=140 \text{ps}}$) and NANF ($\rho_{\text{NANF}, \Delta t=140 \text{ps}}$) in Fig. \ref{fig::concurrence_purity_VS_dt}b and \ref{fig::concurrence_purity_VS_dt}c, respectively. In Fig. \ref{fig::concurrence_purity_VS_dt}a, the dashed and solid lines represent our simulation data for the concurrence and purity of the distributed state. 
As presented in the Appendix, our model incorporates the effect of overlapping time-bins as error counts at the detectors.
The slight disagreement between our model and data in the transition region arises from our use of a simple Gaussian to describe the temporal shape of our pulses.
Fig. \ref{fig::concurrence_purity_VS_dt}a shows that when $\Delta t \to 0$ for both SMF28 and NANF, the concurrence of the distributed quantum state is lost. However, even for $\Delta t \to 0$, some coherence is still preserved and the state purity $\gamma_{\text{SMF28}}$ and $\gamma_{\text{NANF}}$ remain greater than the minimum purity $\gamma_{\text{min}}=\frac{1}{4}$ for a two-qubit maximally mixed state. This is also reproduced by our model.

\section{Discussion}
\label{sec:discussion}
We have presented the distribution of entanglement over a long ($7.7$ km) NANF. To achieve this, we generated a {$\ket{\Psi^-}$~-~Bell state} between a polarization qubit encoded in one photon and a time-bin qubit in anther, transmitting the time-bin qubit through the HCF. We verified our  entanglement distribution by performing two-qubit state tomography and reconstructing the resulting density matrix. This allowed us to quantify the concurrence and purity of the quantum state, finding that the concurrence (purity) decreased slightly from $0.9482\pm0.0007$ ($0.9493\pm0.0008$) to $0.901\pm0.006$ ($0.875\pm0.006$) due to a slight depolarization effect in the NANF. For comparison, we performed entanglement distribution using a 7.8 km SMF28 fiber. We found that, for larger time-bin spacings, the SMF28 fiber outperformed the NANF fiber; i.e. we found no measurable decrease in either the concurrence or purity in this case.
However, when we repeated the experiment with different time-bin spacings, we found the NANF preserves the entanglement for much smaller time spacings, because of its low chromatic dispersion. In particular, we found that in our SMF28 fiber the concurrence already decreases for $\approx~300$~ps spaced time bins, while in NANF the concurrence remains high until $\approx 140$ ps. Moreover, the NANF data decreases primarily due to our detector jitter, rather than dispersion. Although there are techniques to circumvent dispersion in SMF, in NANF low dispersion comes for free. Moreover, NANF has the additional advantages of low optical nonlinearity (which substantially suppresses Raman scattering, allowing strong classical signals to copropagate with quantum light \cite{Honz_co-propagationHCF,Alia2021Coexistence}), and group velocities near $c$ (for ultra-low latency communication). Given the rapid advancements in high-quality AR-HCF fabrication at a variety of operating wavelengths, our work sets the stage for AR-HCF-based quantum communication protocols and quantum photonic technologies.

\section*{Funding}
Horizon Europe research and innovation programme (101071779, 101114043); Engineering and Physical Sciences Research Council (EPSRC) Airguide Photonics project (EP/P030181/1); Horizon 2020 Framework Programme (820474); Austrian Science Fund (10.55776/F71, 10.55776/FG5, 10.55776/I5656); Air Force Office of Scientific Research (AFOSR)(FA9550-21-1-0355); Austrian Federal Ministry for Digital and Economic Affairs; National Foundation for Research, Technology and Development; Christian Doppler Forschungsgesellschaft (501100006012); Erwin Schrödinger Center for Quantum Science and Technology.

\section*{Acknowledgements}
We thank Lennart Jehle and Michal Vyvlecka for use of their low-jitter SNSPD system, Christopher Hilweg for fruitful discussion on AR-HCF applications, and Obada Alia and George T. Kanellos for their assistance with the initial characterization of the hollow-core fiber. This work benefited from network activities through the INAQT network, supported by the Engineering and Physical Sciences Research Council. This research was funded in whole or in part by the Austrian Science Fund (FWF) [10.55776/F71, 10.55776/FG5, 10.55776/I5656].

\section*{Data Availability}
{All the data that are necessary to replicate, verify, falsify and/or reuse this research is available online at \cite{antesberger_michael_2023_8207772}.}

\bibliography{main}

\appendix
\section{Simulation model}
\begin{figure}[h]
    \centering
    \includegraphics[width=0.95\linewidth]{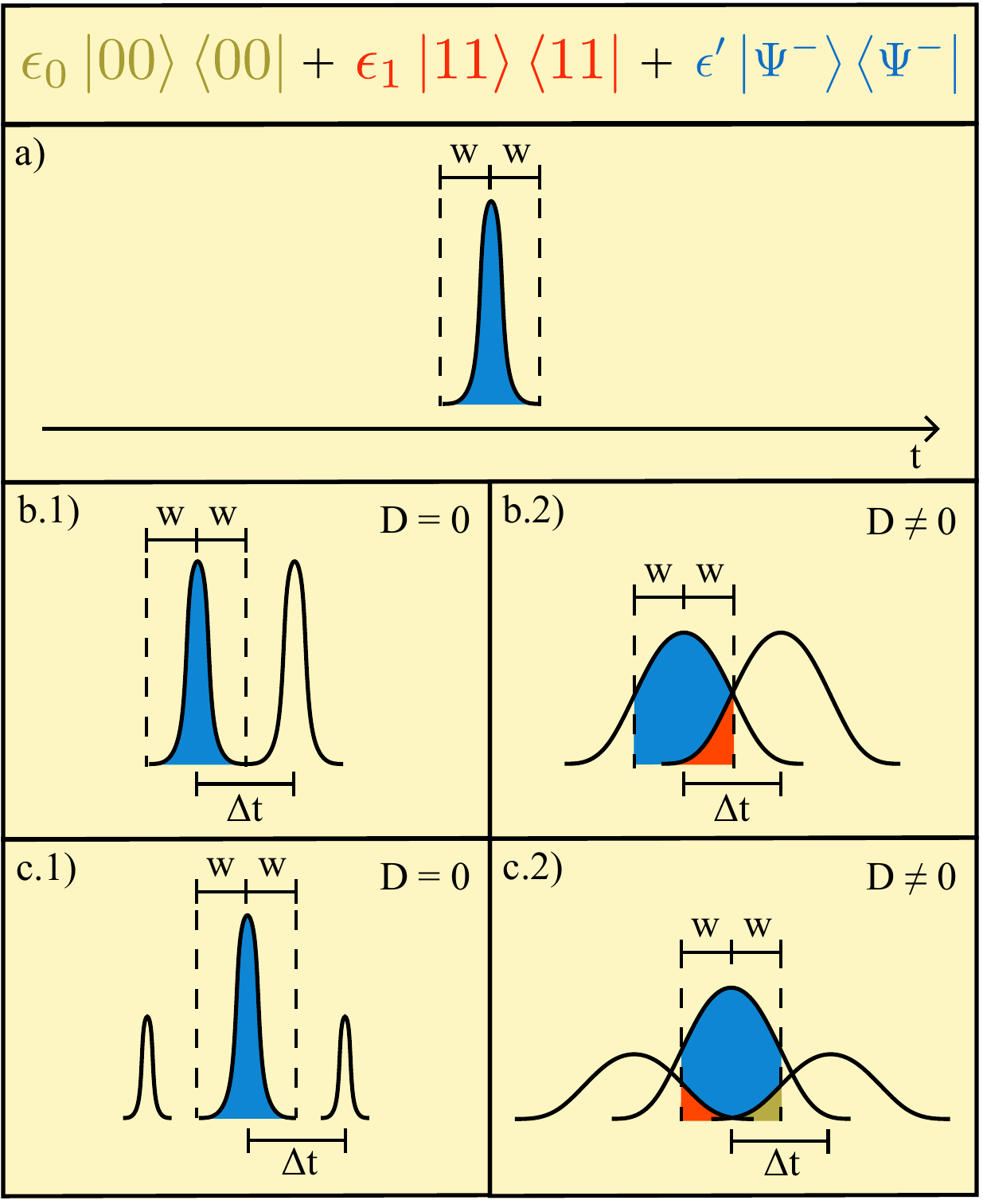}
    \caption{\textbf{Noise Model.} The top row shows Eq. \ref{eq:ideal_model} in colors that correspond to the panels below. \textbf{a)} The temporal distribution of the polarization encoded qubit with the coincidence window $w$. \textbf{b.1)} and \textbf{b.2)} illustrate the wavepacket of the time-bin qubit before and after propagation through a dispersive medium, with the window sown above.  In the presence of dispersion (\textbf{b.1}) the pules overlap and photons in the first time bin can be found in the second. \textbf{c.1)} and \textbf{c.2)} the same scenarios, but after we convert the encoding from time-bin to polarization. Now there are three peaks, and we wish to distinguish the central peak. This cannot be done in the presence of dispersion.}
    \label{fig::model}
\end{figure}

In Section \ref{sec:results} we discussed the dependence of the entanglement quality on the time-bin spacing $\Delta t$. In this measurement we performed QST on the resulting two-qubit state with $\Delta t$ ranging from $0$ to $520$ ps in steps of $20$ ps. As $\Delta t$ drops below $6\sigma$, where $\sigma$ is the standard deviation of the coincidence histogram measured with our detection system (see section \ref{sec:results}), the time-bin modes start to overlap. Consequently, it becomes impossible to properly measure the time-bin qubit. 
Hence, with decreasing time-bin spacings the concurrence and purity approach their minimum values until no entanglement is detectable.
To model this behaviour, we treat cases where the photon ends up at the wrong detector port as error counts, as illustrated in Fig. \ref{fig::model}.

Fig. \ref{fig::model}a shows the temporal distribution of Photon 1. The coincidence window $w$, is chosen to include the entire photon distribution.
Fig. \ref{fig::model}b.1 and \ref{fig::model}b.2 show the temporal distribution of Photon 2's time-bin modes before and after propagation through a dispersive medium (e.g. optical fiber), respectively. 
Now assume we would like to perform a coincidence measurement in the computational basis between the two photons. This means we will need to temporally resolve Photon 2, to determine if Photon 2 is in the first or second time bin. 
(Photon~1, the trigger photon, is not transmitted through a long-distance fiber spool, and encodes a polarization qubit.) 
Therefore, to measure a coincidence between the long (second) time bin and the trigger photon, we center our coincidence window over the long time-bin (Fig. \ref{fig::model}b) and record coincidences events between Photon 1 and Photon~2. 
It is clear that after propagation through a sufficiently dispersive medium, the time-bin modes can overlap, and this can now lead to cases wherein the photon was in the short (first) time-bin mode but it is found in the long time-bin.
We refer to this scenario as an error count.
The situation is similar, but more complex for measurements of the time-bin qubit in other bases.

As we discussed in Section \ref{sec:experiment}, we convert the time-bin qubit back to a polarization qubit using passive optic elements. 
It is well-known, that a passive recombination of two time-bins results then in three prominent peaks. A central peak, which corresponds to the coherent recombination of the time-bin modes and two side peaks. The side peaks arise when, for example, the $\ket{S}$ ($\ket{L}$) mode transmits along the short (long) path after the 50:50 BS. If the time-bins overlap in the time domain, this is transferred to an overlap of the two side peaks each with the central peak. See Fig. \ref{fig::model}c.1 and \ref{fig::model}c.2. Following the optical path through the PBS in Fig. \ref{fig::setup}a, one can see, that the long (short) side peak will always be in $\ket{H}$ ($\ket{V}$), while the central peak holds the information of the Bell state $\ket{\Psi^-}=\frac{1}{\sqrt{2}}\ket{HV}_{12}-\ket{VH}_{12}$. 
When we perform now QST on the two-qubit state, we center our coincidence window on the central peak.
However, because of the overlap with the side peaks (which is caused by the initial dispersion, and would also manifest in other active measurement schemes), the resulting state is no longer a pure $\ket{\Psi^-}$ state.

To understand the above situation, consider again, a measurement in the computational basis. 
Say Photon 1 of the Bell state is found in $\ket{H}$. We would then expect Photon 2 to be in $\ket{V}$. However, if Photon 2 could be found now in the horizontally-polarized side peak. Since this bin overlaps the central time bin, there is now a non-zero probability that we will find the coincidence measurement in $\ket{HH}$. The same applies for the case when Photon 1 is measured in $\ket{V}$. There will be cases wherein Photon 2 ends up in the vertical side peak, and thus the registered coincidence event corresponds to $\ket{VV}$. We can then model the resulting state as
\begin{equation}
    \hat{\rho} = \epsilon_0 \ket{HH}\bra{HH} + \epsilon_1 \ket{VV}\bra{VV} + \epsilon^{'} \ket{\Psi^-}\bra{\Psi^-} \\,
    \label{eq:ideal_model}
\end{equation}
where $\epsilon_0$, $\epsilon_1$ and $\epsilon^{'}$ are probabilities to detect the corresponding state and are calculated as
\begin{align}
    \epsilon_0 = \frac{1}{\sqrt{2}\sigma} \int_{-w}^{w} e^{-\frac{(t+\Delta t)^2}{2\sigma^2}} &dt \,,
    \hspace{2mm}\epsilon_1 = \frac{1}{\sqrt{2}\sigma} \int_{-w}^{w} e^{-\frac{(t-\Delta t)^2}{2\sigma^2}} dt\,, \nonumber\\[2.2mm]
    \epsilon^{'} =& \frac{1}{\sqrt{2}\sigma} \int_{-w}^{w} e^{-\frac{t^2}{2\sigma^2}} dt \hspace{1mm}.
    \label{eq:epsilons}
\end{align}
Here, we assume that the photon distribution is Gaussian. Experimentally, we chose the coincidence window as $w=3\sigma$, so as not to discard coincidence events. Hence, if $\Delta t \geq 6\sigma$ the probabilities $\epsilon_0=\epsilon_1=0$ effectively. To consider imperfection in the initial state generation or concurrence degradation due to fiber properties (e.g. the depolarizing effect of our HCF) we use our experimentally measured density matrix $\hat{\rho}_{\Delta t=520 ps}$ as an input to our model. Hence,
\begin{equation}
    \hat{\rho}^{'} = \epsilon_0 \ket{HH}\bra{HH} + \epsilon_1 \ket{VV}\bra{VV} + \epsilon^{'} \hat{\rho}_{\Delta t=520 ps} \\.
    \label{eq:exp_model}
\end{equation}
The simulation data (dashed and solid lines representing the output concurrence and purity) in Fig. \ref{fig::concurrence_purity_VS_dt}a are calculated from the state in Eq. \ref{eq:exp_model}. Our model fits the small and large time-bin spacings very well. The precise shape in the intermediate regime depends strongly on the photon distribution, which we assumed as Gaussian for simplicity. This assumption leads to slight discrepancy between our data and the model in the intermediate regime.

\end{document}